# Photonic sequential logic circuits for stateful intelligence

Jingcheng Li [1,†], Wenkai Zhang[1,†], Jialong Zhang[1], Bo Wu[1], Qichao Ding[3], Hongli Wang[3], Hailong Zhou[1,2,*], Jianji Dong[1,2,*], and Xinliang Zhang[1,2]

[1]*Wuhan National Laboratory for Optoelectronics, Huazhong University of Science and Technology, Wuhan 430074, China*

[2]*Optics Valley Laboratory, 430074 Wuhan, China*

[3]*Hubei Jiufengshan Laboratory, Wuhan, China.*

[†]*These authors contributed equally: Jingcheng Li, Wenkai Zhang.*

*Corresponding author: hailongzhou@mail.hust.edu.cn, jjdong@mail.hust.edu.cn

## Abstract

By virtue of its high energy efficiency and ultra-low latency, optical computing is regarded as a highly promising computing paradigm in the post-Moore era. However, constructing a Turing-complete optical computing for next-generation AI paradigms, from large language models (LLMs) to real-time autonomous systems, requires not only stateless function mapping units, but also sequential logic circuits capable of storing historical states and processing dynamic data streams. Current photonic processors lack this temporal memory, and the state-of-the-art optical sequential schemes remain impractical due to signal attenuation and the absence of programmable clock control. Here, we propose a photonic sequential logic circuits (PSLC) to resolve this fundamental challenge. Leveraging a uniquely designed active optoelectronic feedback network, we construct a complete family of photonic sequential logic devices, comprehensively encompassing set-reset latches, D latches, and edge-triggered master-slave D flip-flops. We further construct photonic sequence detectors and asynchronous counters, verifying the capability of the architecture to execute both synchronous and asynchronous sequential tasks, alongside its system-level scalability. Ultimately, by combining the constructed PSLC with stateless function mapping units, we establish a universal hardware paradigm for stateful intelligence. We validate this paradigm by executing highly reliable real-time drone obstacle avoidance in the physical domain, alongside Shakespearean-style text generation in the symbolic domain. This work provides the crucial missing piece of the puzzle for optical computing to achieve stateful computing, establishing a definitive hardware foundation for advancing toward Turing-complete optical computing.

Keywords: Optical digital computing, Photonic sequential logic circuits, Photonic computing

## Introduction

In the era of advanced artificial intelligence, the core focus of computation increasingly relies on processing dynamic, sequential data streams. This demand spans a broad spectrum: from autoregressive generation in large language models (LLMs) to real-time, on-device decision-making within the Edge Computing[1]. Consequently, executing these sequence-dependent tasks with ultra-low latency and high energy efficiency has emerged as the primary challenge for next-generation photonic intelligent systems[2]. Despite significant advancements in electronic integrated circuits, critical metrics such as energy efficiency and latency are rapidly hitting their fundamental physical limits[3-6]. These bottlenecks, inherent in traditional electronic hardware architectures, have driven the exploration of new computing architectures designed specifically to overcome these sequential processing limitations[7].

By virtue of its high energy efficiency and ultra-low latency, optical computing is widely regarded as a highly promising computing paradigm to overcome these electronic bottlenecks. Over the past decade, optical computing hardware has experienced explosive growth, particularly in analog accelerators represented by optical matrix multiplication[8-20], as well as digital photonic combinational logic devices[21-27]. Although these feedforward architectures exhibit tremendous advantages in executing offline tasks[28-30] (Fig. 1a(i)) and stateless function mapping relying exclusively on current inputs[31-33] (Fig. 1a(ii)), they complete only half of the puzzle for Turing-complete optical computing. To effectively process the aforementioned dynamic, sequential data streams, the construction of next-generation photonic intelligent systems relies on two indispensable hardware pillars: feedforward architectures for stateless computation, known as combinational logic in digital circuits, and sequential logic circuits responsible for historical state retention and temporal synchronization. As illustrated in Fig. 1a(iii), within real-time multi-time-step processing tasks[34,35] such as LLMs, and drone obstacle avoidance (Fig. 1a(iv)), systems must rely on sequential logic to execute spatio-temporal reasoning[36]. However, photonic sequential logic, representing the missing half of Turing completeness, has lagged significantly. State-of-the-art solutions are constrained by a fundamental hardware trade-off: schemes based on all-optical nonlinearity offer high speeds but are hindered by severe power attenuation, making large-scale cascading impractical[37]. Conversely, approaches relying on fiber or waveguide delay lines exhibit low loss, but their clock frequencies are rigidly locked by physical lengths. Lacking programmability, these delay-line act as fixed analog memories rather than true digital logic[38]. Currently, the field of integrated photonics lacks a sequential logic paradigm capable of simultaneously overcoming this hardware trilemma: achieving monolithic integration, fully controllable clocking, and large-scale cascading potential. It is precisely the absence of such integrated photonic sequential logic circuits that severely restricts the progression of optical computing from static perception toward mastering dynamic, sequential data streams in real-world environments.

To address this long-standing absence, here we propose photonic sequential logic circuits (PSLC) that overcomes the aforementioned hardware trilemma by simultaneously achieving monolithic integration, fully controllable clocking, and large-scale cascading potential (Fig. 1b). By designing and experimentally validating an integrated active optoelectronic feedback network, we realize digital photonic sequential devices featuring robust bistability and active signal regeneration and programmable clock control, encompassing set-reset latches, D latches, and D flip-flops. Building upon these fundamental units, we construct a photonic sequence detector and an asynchronous counter, comprehensively validating the system-level scalability of the architecture in executing both synchronous and asynchronous sequential tasks. Crucially, we demonstrate the seamless system-level combination of this sequential architecture and stateless function mapping units to establish a universal hardware paradigm for stateful intelligence. In the physical domain, an end-to-end demonstration of a real-time drone obstacle avoidance task achieves a 97.7% navigation success rate in complex environments, delivering a substantial 56.8% absolute increase in success rate compared to purely feedforward photonic systems lacking temporal memory. Concurrently, in the symbolic domain, we implement a photonic generative language model to validate the cascaded depth of our architecture. By exploiting a large-scale PSLC array for extensive historical context retention and physical position embedding, the system successfully executes stateful, character-level autoregressive generation, emulating Shakespearean text with over 85% spelling accuracy and 70% style fidelity. Our work not only highlights the unique advantages of this system in executing complex spatio-temporal reasoning and cognitive sequence generation tasks, but also marks a significant milestone in the evolution of optical computing toward Turing-complete hardware for next-generation photonic intelligent systems.

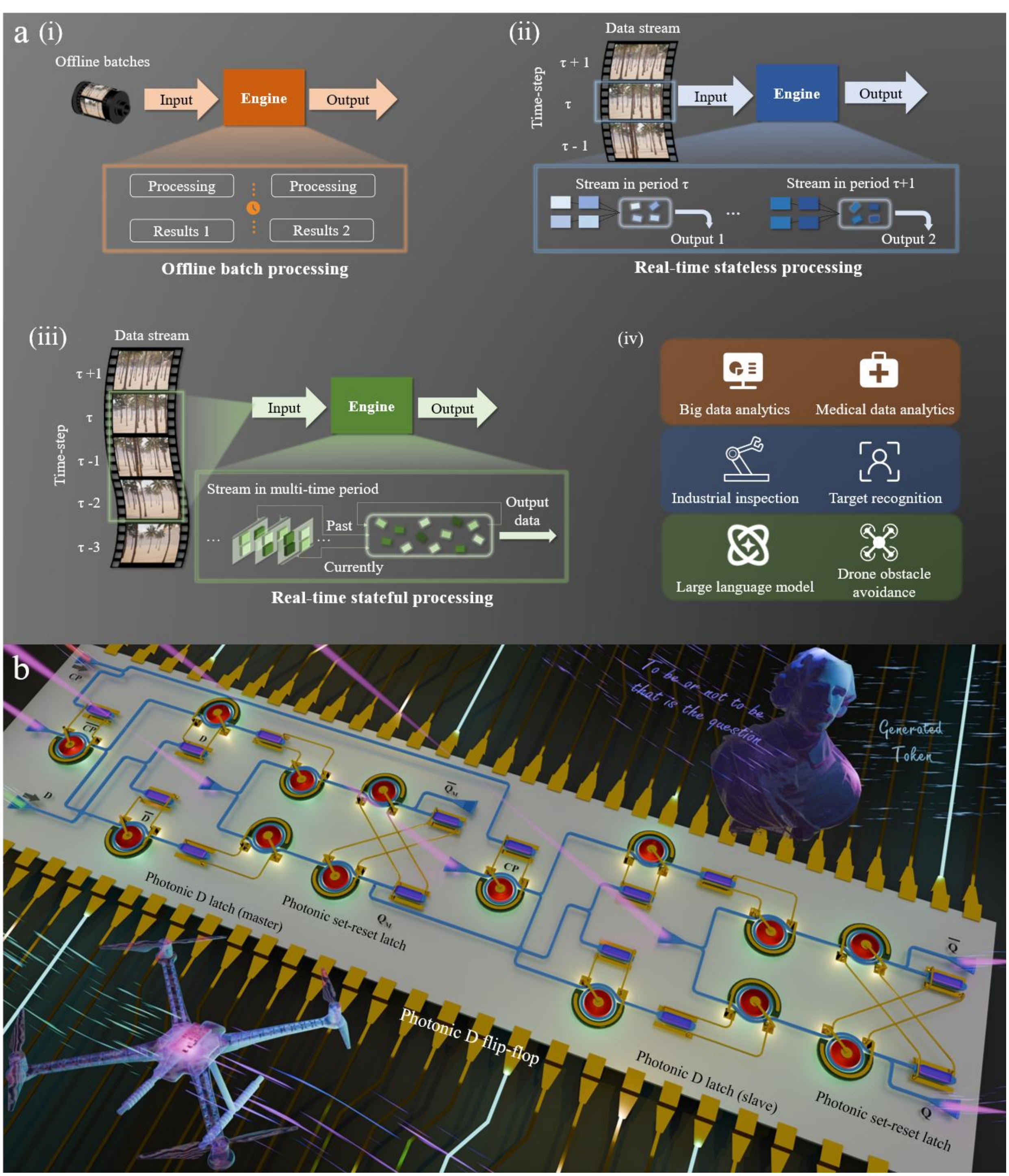

Fig. 1 Categories of computing paradigms and the photonic sequential logic architecture. (a) Three computing paradigms classified by temporal dependency and their typical application scenarios. (i) Offline batch processing ,which primarily handles offline static datasets. (ii) Real-time stateless processing, where the computational output depends exclusively on the current input slice, rendering it completely stateless. In contrast, (iii) real-time stateful processing necessitates sequential logic circuits to execute

complex spatio-temporal reasoning tasks, where decision-making relies on the deep synergy between current inputs and historical states. (iv) Typical application scenarios of three computing paradigms. (b) Schematic of the proposed chip-scale PSLC.

## Results

### Photonic D flip-flop

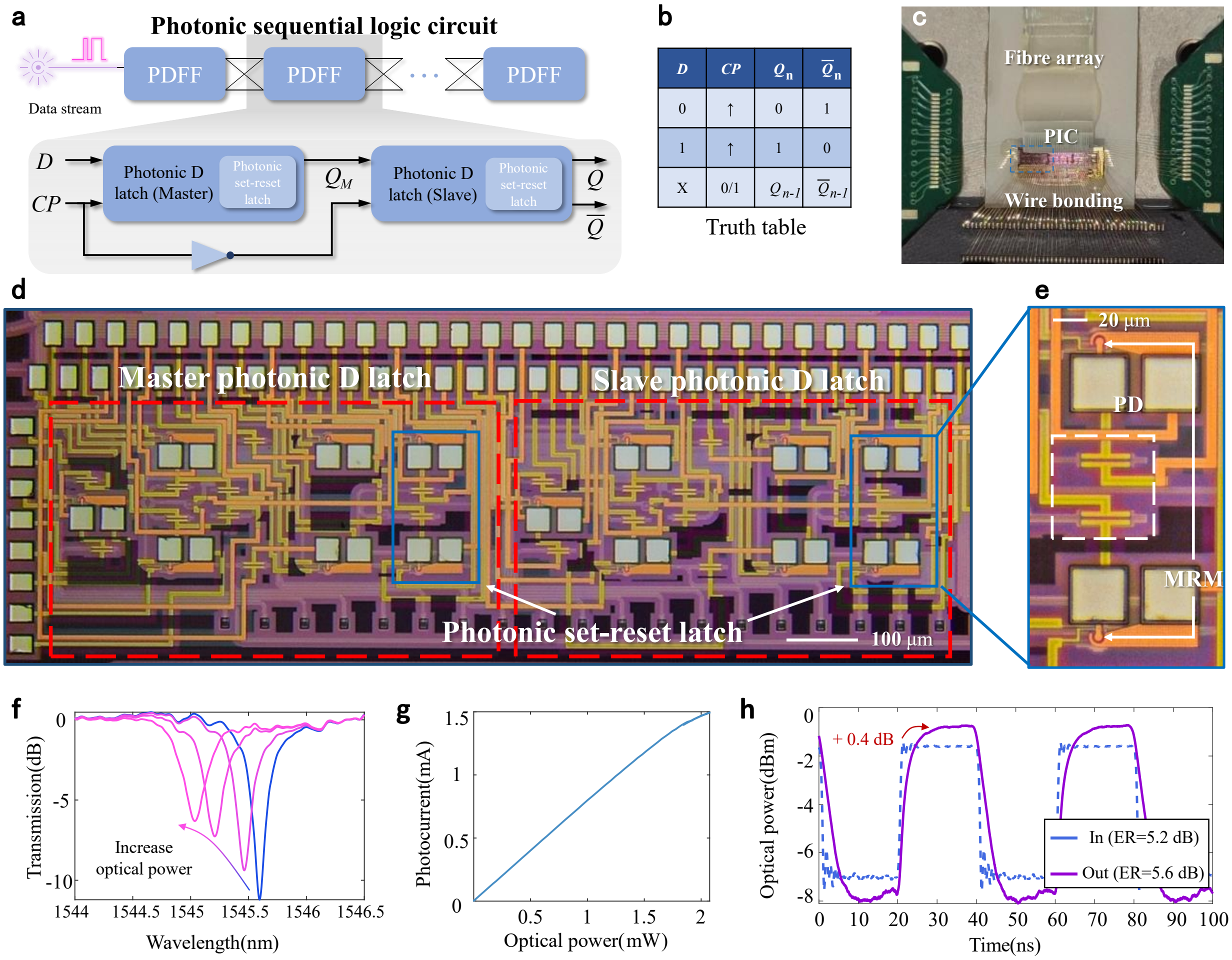

| $D$ | $CP$ | $Q_n$ | $\overline{Q}_n$ |
|---|---|---|---|
| 0 | ↑ | 0 | 1 |
| 1 | ↑ | 1 | 0 |
| X | 0/1 | $Q_{n-1}$ | $\overline{Q}_{n-1}$ |

Fig. 2 Principle and chip characterization of the photonic D flip-flop (PDFF). (a) Schematic of the edge-triggered photonic master-slave D flip-flop. (b) Truth table. (c) Photograph of the packaged device illustrating the integration of the fibre array, photonic chip, and wire bonding. (d) Micrograph revealing the physical layout of a single PDFF. (e) Zoomed-in micrograph of the cross-coupled PD-MRMs. (f) Measured resonance shift of the microring transmission spectrum. (g) Photocurrent response of the PD under a 3 V forward bias. (h) Measured dynamic input and output optical signals of the PD-MRM. PD, photodetector; MRM, microring resonator modulator.

To bridge the hardware gap for processing dynamic, sequential data streams, we first present the integrated photonic master-slave D flip-flop—the cornerstone of modern sequential logic circuits requiring rigorous clock-edge synchronization (Fig. 2a). Analogous to its electrical counterpart, this architecture comprises two cascaded photonic D latches driven by optical clocks with identical frequencies but strictly inverted phases. This complementary clock-driving mechanism effectively eliminates the transparent window inherent in a single latch. As depicted by the truth table in Fig. 2b, the device is strictly positive-edge-triggered. The output terminal precisely captures the state of the data input *D* exclusively at the rising edge of the clock signal *CP*. At all other times, the input and output remain isolated, ensuring the output sustains a robust steady state. This nanosecond-scale temporal discretization serves as the critical hardware prerequisite for realizing complex pipelining and real-time spatio-temporal reasoning. Physically, this chip overcomes the limitations of traditional passive optical computing architectures by adopting an active integrated design. Figs. 2c and 2d present a photograph of the packaged chip and an optical micrograph of a single PDFF cell, respectively. Furthermore, Fig. 2e details the cross-coupled photodetector-driven microring modulators (PD-MRM), which structurally instantiate the active optoelectronic feedback network at the fundamental physical core of this sequential logic engine.

To validate the performance of this fundamental unit, we conducted comprehensive physical characterizations (Figs. 2f–h). As the input optical power increases, the nonlinear electro-optic effect within the MRM drives a pronounced resonance shift in the transmission spectrum. Concurrently, under a 3 V forward bias, the PD exhibits a robust photocurrent response, achieving a responsivity of 0.8 A/W. Crucially, a comparison of the dynamic waveforms reveals that the extinction ratio (ER) of the output signal does not degrade; instead, it is enhanced by 0.4 dB relative to the input. This result demonstrates that the PD-MRM unit possesses intrinsic optical signal regeneration capabilities. By actively compensating for the insertion and propagation losses of the optical link, this built-in regeneration overcomes the severe attenuation bottlenecks inherent in traditional optical computing architectures. Ultimately, it endows the photonic sequential logic circuits with the essential capability for large-scale cascading and large-scale system scalability.

## Photonic set-reset latch

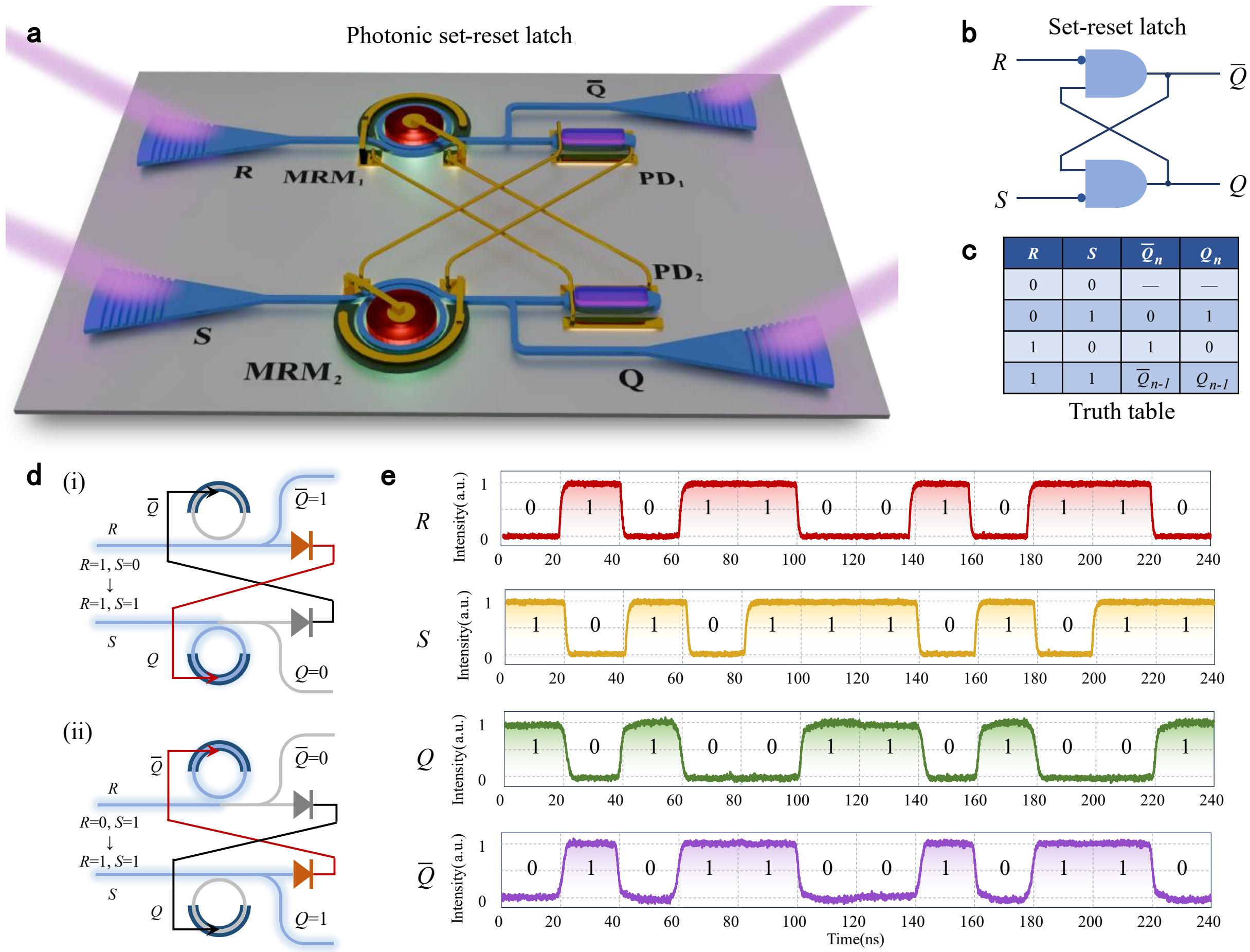

Fig. 3 Principle of the photonic set-reset latch. (a) Physical architecture. (b) Equivalent logic gate circuits. Solid dots and "D"-shaped symbols denote logical NOT and AND operations, respectively. (c) Truth table. (d) Transient response analysis showing the device transitions: (i) from the reset mode to the hold state, robustly maintaining its prior reset level, and (ii) from the set mode to the hold state, sustaining its prior set level. (e) Dynamic input and output waveforms. MRM, microring resonator modulator; PD, photodetector; CW, continuous wave.

Within the PDFF architecture, the fundamental capacity to capture and retain historical states originates directly from its core physical building block: the integrated photonic set-reset latch—which constitutes our proposed active optoelectronic feedback network. This foundational unit is governed by the characteristic equation

$$Q_n = \overline{R} + SQ_{n-1} \quad (1)$$

Fig. 3a illustrates the physical architecture of the photonic set-reset latch, comprising two cross-coupled PD-MRMs. As established, the system exhibits robust bistability under specific optical input conditions, thereby providing the physical substrate for optical signal memory. At the system level, we directly map these complex physical dynamics onto a standard digital circuits architecture (Fig. 3b). Within this framework, R and S are designated as the reset and set optical signal input ports, respectively, while the system generates a pair of strictly complementary logic outputs, $Q$ and $\overline{Q}$. The static logic response of the device rigorously conforms to the constraints of standard digital sequential circuits. Its truth table, depicted in Fig. 3c, details the four fundamental operational modes:

- Hold state ($R$=1 and $S$=1): The device utilizes the active optoelectronic feedback network to robustly maintain the output state of the preceding cycle ($Q_n = Q_{n-1}$). This mechanism provides the fundamental physical basis for memory within the optical domain.
- Set state ($R$=0 and $S$=1): The input configuration overrides the existing equilibrium of the feedback network, driving the output terminal to a stable logic-high state ($Q_n = 1$).
- Reset state ($R$=1 and $S$=0): Conversely, this input configuration forces the output to transition to a stable logic-low state ($Q_n = 0$).
- Forbidden state ($R$=0 and $S$=0): The absence of valid optical inputs at both terminals fails to provide sufficient optical power to sustain the active feedback loop, rendering the device logically inactive.

To evaluate the practical operational dynamics of this foundational unit, we conducted comprehensive transient and dynamic high-speed experiments. Fig. 3d(i) details the transient response of the device as it switches from the reset mode to the hold state. In the initial reset mode, the absence of optical input at the S port ($S$=0) yields a zero photocurrent from the corresponding lower PD. Consequently, the upper MRM remains in an off-resonance state. This configuration allows the optical power of signal R to propagate with negligible loss, directly reaching the output terminal $\overline{Q}$ while simultaneously driving the upper PD. Upon switching to the hold state ($R$=1, $S$=1), the strong photocurrent generated by the upper PD directly drives the lower MRM, forcing it into an on-resonance state. This active optoelectronic feedback effectively inhibits the transmission of signal $S$ to the $Q$ terminal, thereby ensuring that the lower output is physically locked at a logic-low level ($Q$=0). Conversely, Fig. 3d(ii) illustrates the symmetric transient process transitioning from the set state to the hold state. Driven by the mirror-complementary feedback path, this mechanism robustly ensures that the output terminal is firmly latched at a logic-high level ($Q$=1). These two representative sets of transient measurements collectively verify the robust bistability of our proposed active optoelectronic feedback network. Building upon this physical foundation, we conducted dynamic measurements to validate the capability of the device to process dynamic, sequential data streams (Fig. 3e). Operating at 50 MHz under aperiodic $R$ and $S$ input signals, the generated $Q$ and $\overline{Q}$ sequences exhibit strict agreement with the designed Boolean truth table. The steep, nanosecond-scale state transitions not only demonstrate the dynamic noise immunity of the device but also signify its physical readiness as the core hardware primitive for constructing the aforementioned edge-triggered PDFF and scaling toward complex spatio-temporal reasoning architectures.

## Photonic D latch

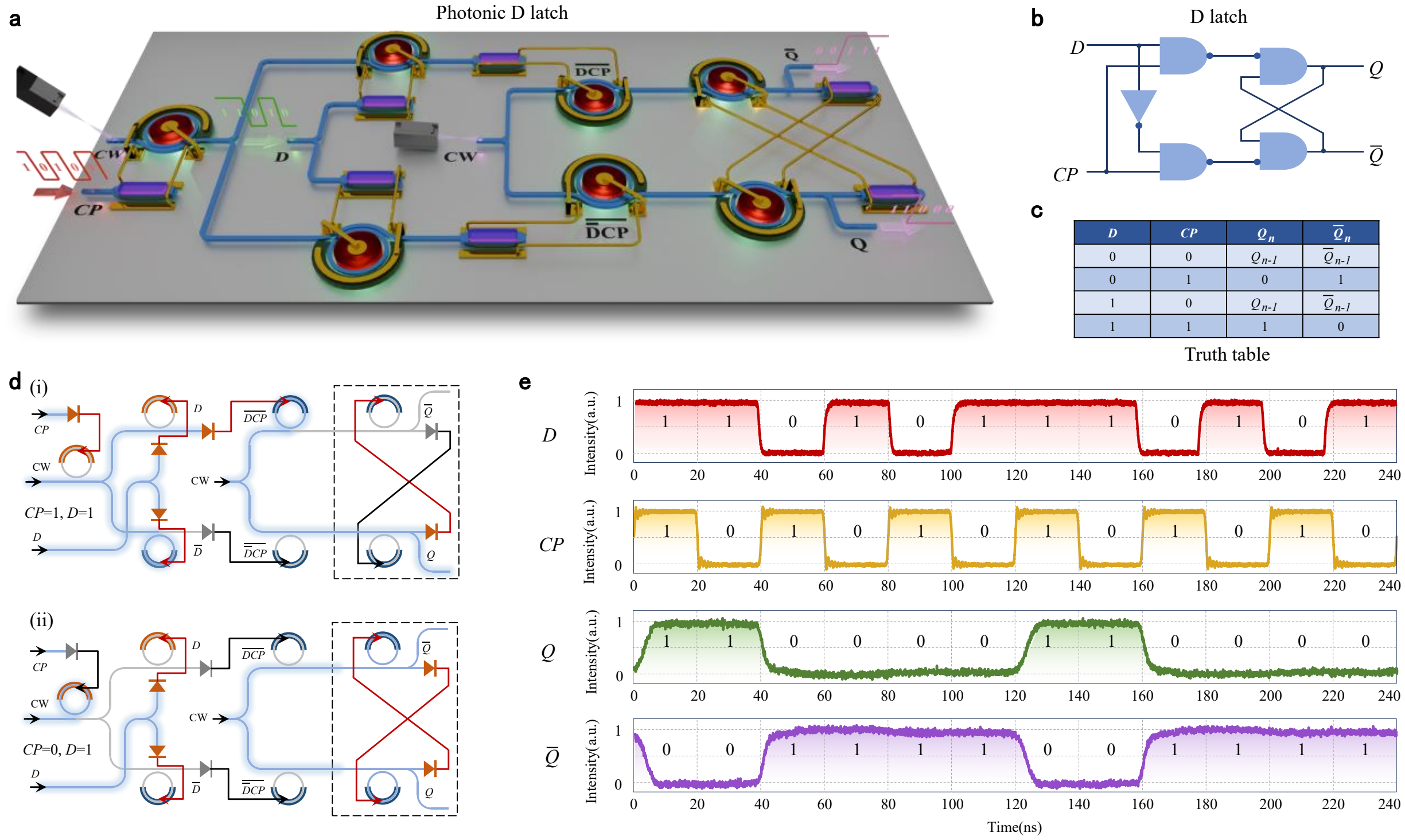

| $D$ | $CP$ | $Q_n$ | $\overline{Q}_n$ |
|---|---|---|---|
| 0 | 0 | $Q_{n-1}$ | $\overline{Q}_{n-1}$ |
| 0 | 1 | 0 | 1 |
| 1 | 0 | $Q_{n-1}$ | $\overline{Q}_{n-1}$ |
| 1 | 1 | 1 | 0 |

Fig. 4 Principle of the photonic D latch. (a) Physical architecture. (b) Equivalent logic gate circuits. (c) Truth table. (d) Static response analysis: (i) under the transparent mode (*CP*=1), output *Q* strictly tracks input *D*; (ii) under the hold mode (*CP*=0), the output *Q* robustly maintains its previous logic level. (e) Dynamic input and output waveforms.

To advance from the asynchronous memory of the set-reset latch to the fully controllable clocking required by the PDFF, we integrate a front-end clock-gating mechanism, evolving the architecture into a synchronous photonic D latch. This architectural extension achieves precise temporal control over data write operations, governed by the characteristic equation

$$Q_n = D \cdot CP + Q_{n-1} \cdot \overline{CP} \tag{2}$$

Fig. 4a illustrates the physical architecture of this photonic D latch. It comprises a front-end control network constructed from five PD-MRMs interconnected via a specific optical routing scheme, which is subsequently cascaded with the aforementioned photonic set-reset latch. Driven by the data signal *D* and the clock signal *CP* as inputs, the system synchronously generates a pair of strictly complementary logic outputs, *Q* and $\overline{Q}$. At the system level, this complex photonic topology maps directly onto the standard digital logic architecture depicted in Fig. 4b, incorporating three NOT gates, two NAND gates, and two AND gates. Crucially, this gating network successfully introduces the programmable clock control lacking in traditional optical delay lines. Its static logic response is illustrated by the truth table in Fig. 4c, defining two fundamental operational modes:

- Hold mode (*CP*=0): As indicated in rows 1 and 3 of the truth table, the front-end gating network physically blocks any signal fluctuations from the data terminal *D*, ensuring the output remains strictly locked to the state stored by the core RS latch from the preceding cycle.
- Transparent mode (*CP*=1): As indicated in rows 2 and 4 of the truth table, the front-end optical pathways are unblocked, allowing the logic level of the output terminal *Q* to precisely track the real-time variations of the data input *D*.

To validate the robustness of this synchronous architecture, we conducted comprehensive system-level characterizations of the photonic D latch. Fig. 4d presents the static optical power distribution under two representative sets of steady-state inputs, while the dynamic response waveforms in Fig. 4e capture the temporal evolution of the device driven by continuous data streams. These results clearly demonstrate that the output terminal responds to the signal *D* exclusively when *CP* is at a logic-high level, exhibiting robust state retention during all other periods. Crucially, by cascading two such level-sensitive D latches driven by inverted clocks, we successfully eliminate this transparent window, culminating in the strictly edge-triggered PDFF. This completes the fundamental hardware hierarchy required to execute complex spatio-temporal reasoning tasks.

## Demonstration of PSLC's functions

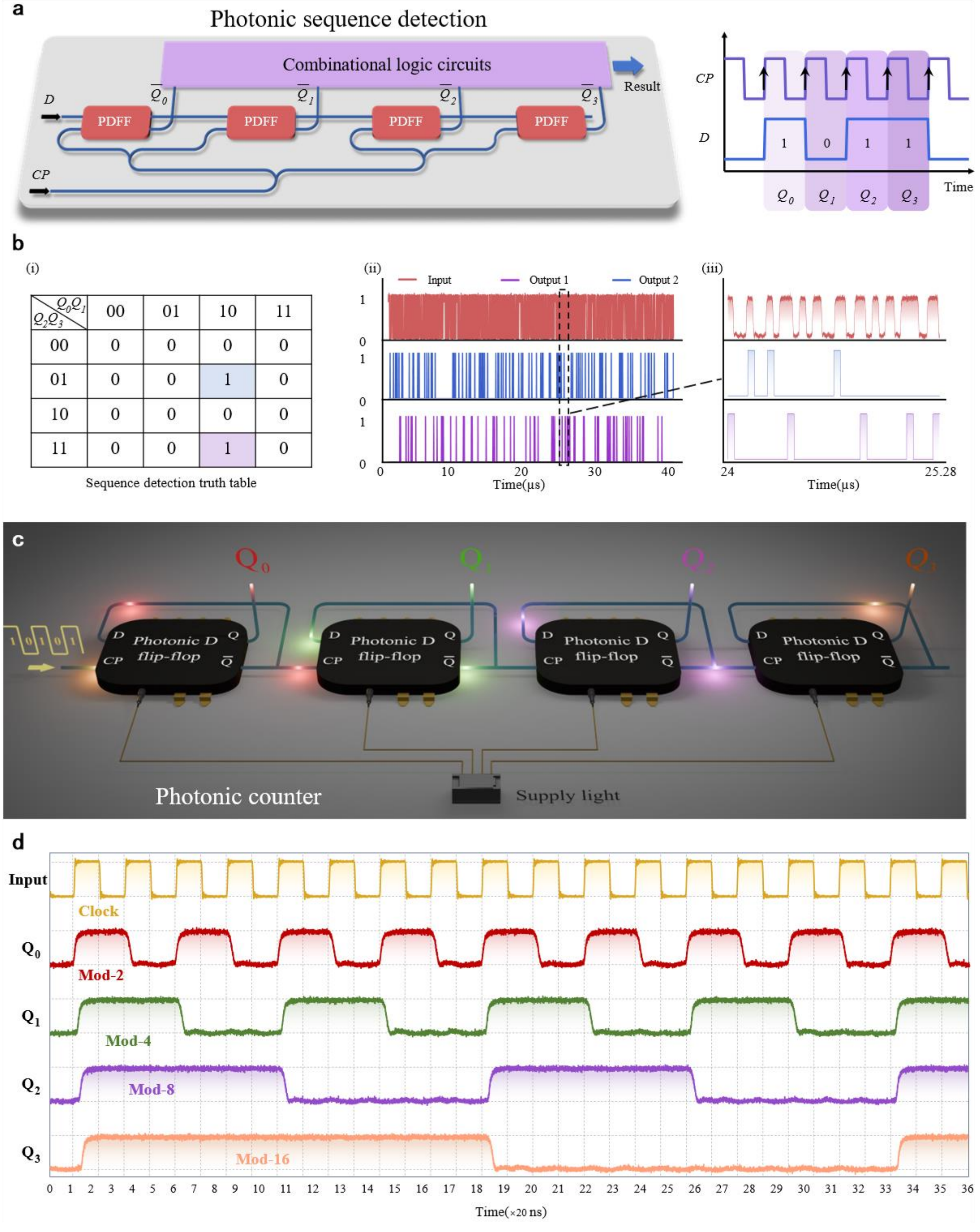

Fig. 5 Construction of synchronous and asynchronous sequential circuits using PDFFs. (a) Schematic of the synchronous 4-bit photonic sequence detector. (b) Experimental characterization. (i) Truth tables for detecting the primary ("1001") and secondary

("1011") target sequences.. (ii) Measured input and output dynamic waveforms of the system. (iii) Magnified view detailing the transient state transitions. (c) Schematic of the 4-bit asynchronous counter. (d) Dynamic temporal waveforms of the counter.

Leveraging the active signal regeneration and robust cascading capabilities of the PDFF, we scale the architecture to construct both synchronous and asynchronous photonic sequential circuits. Fig. 5a illustrates the schematic of a synchronous 4-bit photonic sequence detector. This system comprises a 4-bit photonic shift register—formed by four cascaded PDFFs governed by a single global clock—integrated with back-end combinational logic. Upon each rising clock edge, the input data stream advances by one stage along the register chain. Consequently, within any specific clock cycle, the register outputs the current input alongside the states of the three preceding cycles in parallel. Crucially, this mechanism realizes the physical mapping of serial temporal data into parallel spatial states, directly providing the hardware foundation required for the aforementioned complex spatio-temporal reasoning tasks. Experimental characterizations of this detector are presented in Fig. 5b. Guided by the truth table of the back-end logic (Fig. 5b(i)), the measured dynamic waveforms (Fig. 5b(ii, iii)) demonstrate the successful real-time identification of target data patterns, confirming the architecture's capability for processing dynamic, sequential data streams.

To demonstrate the architectural universality and topological flexibility of this platform, we reconfigured the interconnection strategy to construct a 4-bit photonic asynchronous counter operating without a global clock (Fig. 5c). In this system, an external clock directly drives only the first-stage PDFF. The complementary output of each stage ($\overline{Q_0}$, $\overline{Q_1}$, $\overline{Q_2}$, $\overline{Q_3}$) is fed back into its own data input *D*, while simultaneously serving as the driving clock source for the subsequent stage. As revealed by the dynamic waveforms (Fig. 5d), the application of external clock pulses triggers the first-stage PDFF to execute a modulo-2 counting function. Specifically, for every two transitions of the input clock, its output state toggles exactly once, achieving a precise frequency division by two. By cascading four such stages, the system ultimately realizes a modulo-16 counting task. The successful demonstration of these system-level prototypes clearly validates the robustness and logical completeness of the proposed PSLC, establishing the indispensable temporal memory required for Turing-complete optical computing.

## PSLC for real-time dynamic vision tasks

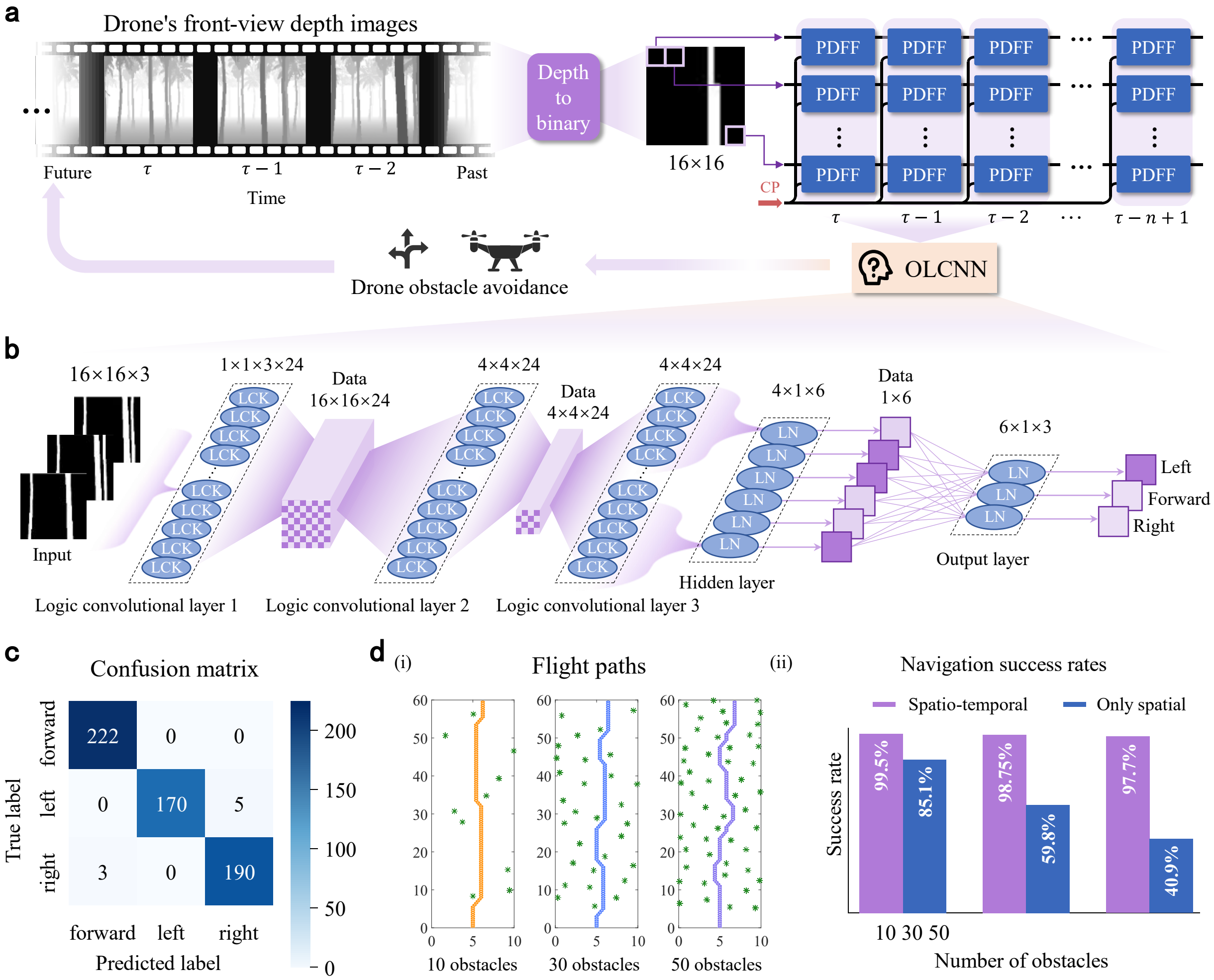

Fig. 6 Application of the PSLC in dynamic real-time tasks. (a) System architecture for drone obstacle avoidance. (b) Schematic of the OLCNN. (c) Confusion matrix for obstacle avoidance classification. (d) Simulated validation of real-time flight performance. (i) Representative flight trajectories in sparse, moderate, and highly dense obstacle environments. (ii) Navigation success rates across the three environmental densities. The performance of the PSLC-driven network (utilizing three consecutive data frames) is compared against a purely combinational logic baseline (restricted to a single data frame).

To validate the system-level efficacy of the PSLC in executing complex real-world dynamic tasks, we target real-time drone obstacle avoidance, a classical spatio-temporal reasoning task (Fig. 6a). Unlike static image classification, dynamic obstacle avoidance relies not only on the current environmental snapshot but also demands the integration of historical states. This integration is crucial for extracting the relative displacements and motion trajectories of obstacles to formulate predictive decisions. To achieve this, we construct a photonic sequential sensing array capable of receiving high-dimensional dynamic data in parallel. This array features a two-dimensional shift buffer network built from large-scale cascaded PDFFs. Within this architecture, a single column latches the spatial pixels of a single frame, while a single row records the temporal evolution of a fixed spatial location across successive time steps. After undergoing binarization preprocessing (where obstacles are labeled as 1 and free space as 0), the drone's front-view depth images are seamlessly fed into this photonic sequential array. This process provides a strictly aligned spatio-temporal data stream for subsequent combinational logic reasoning. To process these streams, we design an optical logic convolutional neural network[39] (OLCNN), with its topological architecture illustrated in Fig. 6b. The input layer of this network simultaneously processes three consecutive frames of spatio-temporal binary images, each with dimensions of 16×16 pixels. The feedforward process is governed by three sets of LCKs. The first layer utilizes a spatio-temporal kernel of size 1×1×3×24 to fuse three frames along the temporal axis, mapping the input into a 16×16×24 feature tensor. Subsequently, the data passes through two successive 4×4 spatial logic convolutions for dimensionality reduction and feature abstraction, yielding a 24-element high-dimensional feature vector. These features are then flattened and fed into a fully connected network comprising two layers of LNs, which ultimately outputs three discrete obstacle avoidance commands (fly left, fly forward, and fly right). Classification testing validates the robust accuracy of this architecture. As depicted by the confusion matrix in Fig. 6c, the system achieves an overall classification accuracy of 98.64% across a large dataset of dynamic obstacle avoidance samples.

To rigorously validate the system's practical utility, we deploy the trained optical logic neural network into a simulated drone flight controller for end-to-end testing in a complex three-dimensional environment (Fig. 6d). Across sparse (10 obstacles), moderate (30 obstacles), and highly dense (50 obstacles) environments, the drone driven by the photonic sequential logic neural network consistently generates successful obstacle avoidance trajectories (Fig. 6d(i)). Crucially, we conduct a controlled evaluation comparing this sequential system (integrating three consecutive spatio-temporal frames) against a purely feedforward baseline (restricted to a single spatial frame) (Fig. 6d(ii)). The experimental data reveal a significant performance divergence. In sparse environments, the performance gap between the two architectures remains marginal. However, as the environment complexity escalates, the purely feedforward logic system, lacking temporal memory, fails to anticipate dynamic motion trajectories. Consequently, its performance degrades significantly, dropping to a 40.9% success rate in highly dense scenarios. Conversely, by accurately capturing the dynamic, sequential data streams, the photonic sequential system robustly sustains a 97.7% success rate, delivering a substantial 56.8% absolute increase compared to the feedforward baseline. This firmly confirms the capability of our proposed PSLC-driven architecture to process dynamic, sequential data streams, successfully realizing the complex spatio-temporal reasoning required for next-generation photonic intelligent systems.

## PSLC for generative language models

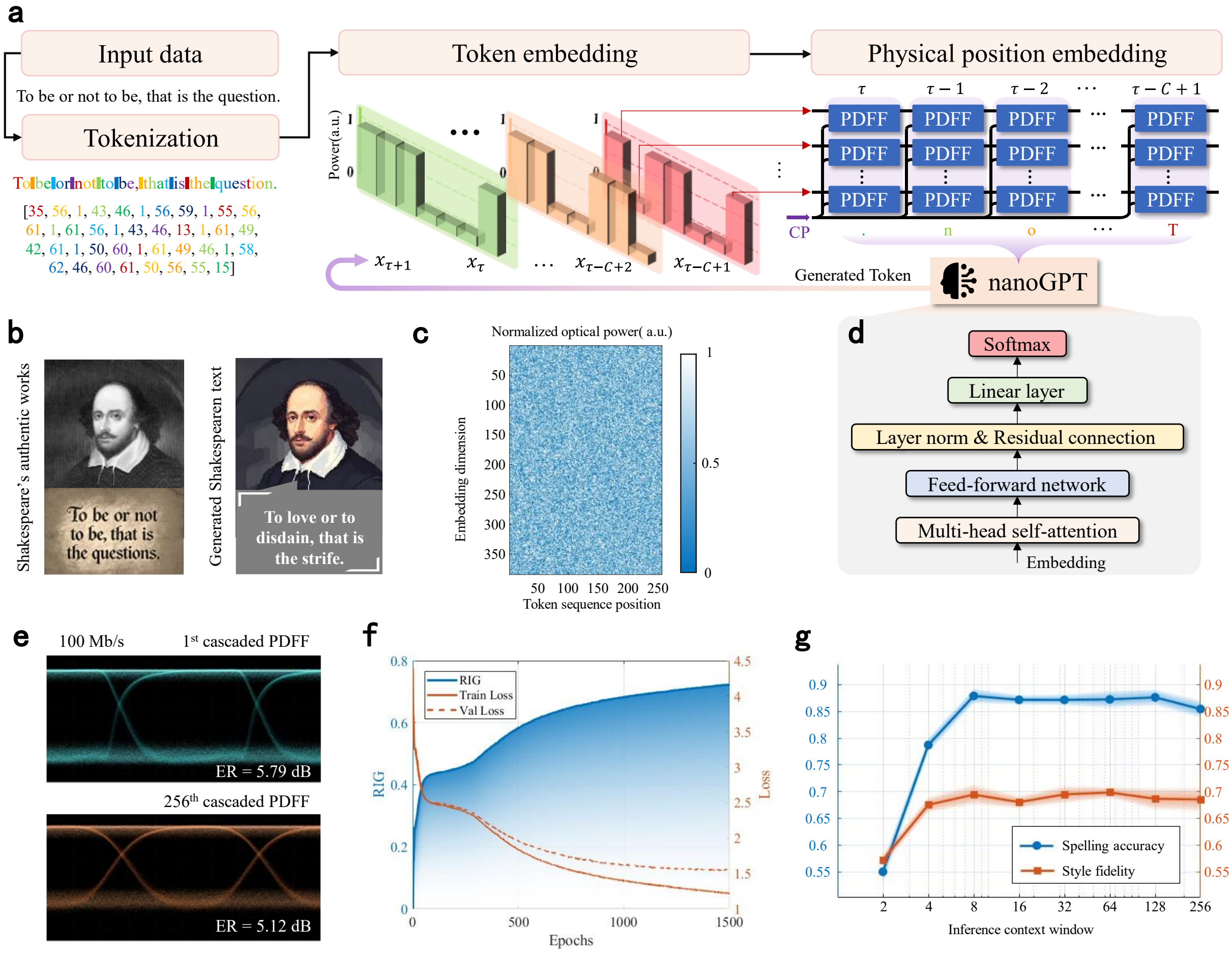

Fig. 7 Application of the PSLC in generative language models. (a) System architecture for character-level Shakespearean text generation. (b) Schematic of the OLCNN. (c) Visualized output states of the 384×256 PDFF array during the character embedding phase. (d) Architecture of the nanoGPT. (e) Eye diagrams of the 1st and 256th cascaded PDFF stages. (f) Training performance showing the convergence of cross-entropy loss and Relative Information Gain (RIG). (g) Quantitative evaluation of spelling accuracy and style fidelity as a function of the inference block size.

To further demonstrate the potential of the PSLC in executing high-order computational tasks, we implemented a generative language model driven by the proposed PDFF array. The autoregressive nature of large language models dictates that the probability distribution of the next token, $P(x_{t+1} \mid x_1, x_2, \ldots, x_t)$, strictly depends on the historical sequence. In our physical architecture, the feedforward network of the digital nanoGPT functions as highly complex state transition mapping. Concurrently, the PDFF array serves as the core sequential memory foundation, responsible for latching and updating the historical context.We selected character-level Shakespearean text generation as the specific demonstration task. As illustrated in Fig. 7a, the input text is first tokenized, where each character is mapped to a specific integer index based on a 65-character vocabulary and embedded into binary representations. Distinct from conventional software algorithms that require explicit computation of positional vectors, our architecture naturally achieves physical position embedding. Because the tokens are sequentially pushed into the PDFF array, the physical stage index of the data within the shift registers inherently encodes its positional context. These states are subsequently fed into the feedforward structure of the digital nanoGPT (Fig. 7d) to predict the next character, thereby forming a complete autoregressive closed loop. Fig. 7b presents the generated samples after training on the complete works of Shakespeare, exhibiting a distinct classical dramatic style.Physical fidelity at the hardware level is essential for supporting the long context windows required by large models. Fig. 7c visualizes the transient power distribution of the character-level embedding signals within the 384×256 PDFF array. The eye diagrams in Fig. 7e further confirm that even after undergoing 256 stages of physical cascading and shifting, the extinction ratio of the signals is perfectly preserved. This validates the extreme robustness of the device architecture in supporting large-scale sequence caching.Furthermore, we quantitatively evaluated the algorithmic performance of the system. As shown in Fig. 7f, the cross-entropy loss and RIG reach stable convergence at 1,500 epochs. Fig. 7g reveals the decisive impact of hardware memory capacity (block size) on the emergence of intelligent behavior: when the inference context window is fewer than 8 characters, the generated text suffers from severe spelling errors and semantic confusion. However, as the window expands beyond 8 characters, the spelling accuracy climbs rapidly and stabilizes above 85% (excluding reasonable divergences caused by obscure archaic English roots and specific character names, the core spelling accuracy exceeds 90%). Simultaneously, the generated text achieves a style fidelity of 70%. This demonstrates that the system has successfully captured and internalized specific high-frequency syntactic distributions from the target corpus, such as dramatic formatting conventions and archaic vocabulary. These results firmly establish the feasibility of photonic sequential circuits in executing complex symbolic sequence generation tasks. These results not only validate the architectural superiority of the proposed PSLC but also mark a definitive milestone in optical computing: the transition from stateless spatial perception toward Turing-complete optical computing for next-generation photonic intelligent systems.

## Discussion

In this work, we propose and experimentally demonstrate a chip-scale PSLC based on PD-MRMs, establishing a complete family of standard photonic sequential logic devices, including photonic D flip-flops, set-reset latches, and D latches. Driven by an active optoelectronic feedback network, this architecture realizes discrete-time state retention, programmable clock control, and digital logic with large-scale cascading potential. Consequently, it overcomes the physical bottlenecks that confine existing photonic processors to stateless feedforward architectures[40,41]. Building upon the PDFF as a core hardware primitive featuring active signal regeneration and large-scale cascading potential, we further scale the architecture into system-level circuits. By designing and experimentally validating photonic sequence detectors and asynchronous counters, we comprehensively verify the logical completeness and topological flexibility of this architecture in executing both synchronous and asynchronous sequential computing tasks. These results not only confirm the reliability of stateful, sequence-dependent processing within the optical domain but also establish a definitive hardware foundation for processing dynamic data patterns.

To evaluate the spatio-temporal reasoning capabilities of this architecture in high-dimensional dynamic scenarios, we apply the system to a real-time drone obstacle avoidance task. In complex environments characterized by dense obstacles, where the integration of multi-frame historical states is imperative for continuous prediction[42,43], the photonic sequential system achieves a 97.7% navigation success rate. This delivers a substantial 56.8% absolute increase in success rate over purely photonic feedforward baselines lacking temporal memory. This system-level demonstration verifies that hardware architectures equipped with temporal memory can efficiently execute ultra-low-latency dynamic spatio-temporal reasoning tasks. Furthermore, to validate the extreme cascading robustness and deep historical context retention of our architecture in the symbolic domain, we implement a photonic generative language model. By utilizing a large-scale PDFF array to inherently provide physical position embedding and stateful temporal memory, the system successfully executes character-level autoregressive generation. Emulating context-dependent Shakespearean text with over 85% spelling accuracy and 70% style fidelity, this demonstration proves that optical temporal memory can reliably support the extensive logic depth required by complex cognitive algorithms. Ultimately, by conquering both continuous physical control and discrete symbolic generation, this work provides the indispensable hardware foundation for optical computing to process complex, real-world dynamic data streams, marking a definitive step toward Turing-complete optical computing for next-generation photonic intelligent systems.

Beyond resolving the immediate bottlenecks of sequential computing, this work establishes a highly scalable foundation for next-generation photonic intelligence. Architecturally, the proposed PDFFs can be seamlessly integrated with combinational logic modules[44], such as emerging optical neural networks or optical tensor cores. This integration facilitates hybrid computing paradigms that unite massive spatial parallelism with complex spatio-temporal reasoning. From an engineering standpoint, the architecture's inherent robustness against manufacturing variations and timing jitter confirms its viability for large-scale integration. Future research can further push the fundamental limits of energy efficiency and clock frequency through the optimization of underlying materials and layout designs, ultimately driving the optical-electrical-optical (OEO) bandwidth to 10 GHz[45]. Concurrently, the introduction of wavelength-division multiplexing[46] (WDM) or multi-mode[47] channel parallel operations will provide the system with immense concurrency, enabling the processing of massive, dynamic data streams. Ultimately, this monolithically integrated digital sequential logic paradigm is poised to propel optical computing from traditional stateless spatial perception toward Turing-complete, next-generation real-time intelligent systems—capable of empowering autonomous navigation, foundational robotic reflex control, and adaptive edge computing.